\documentclass[preprint,epsf]{revtex4}
\usepackage{amsmath}
\usepackage{epsfig}
\usepackage{epstopdf}
\usepackage{dcolumn}
\usepackage{booktabs}

\begin{document}

\title{Dynamical Critical Properties of the Random-Bond three-state Potts Model}

\author{Zeynep Demir Vatansever$^1$}
\author{Ulvi Kanbur$^2$}
\author{Mohammad Sadra Najafi$^3$}
\author{Muktish Acharyya$^4$}
\author{Erol Vatansever$^1$}

\affiliation{$^1$Department of Physics, Dokuz Eyl\"{u}l
University, TR-35160, Izmir-Turkey}
\affiliation{$^2$Department of Physics, Karabük University, Demir Çelik Campus, 78050 Karabük, Turkey}
\affiliation{$^3$ Department of Electrical and Electronics Engineering, Dokuz Eylül University, TR-35160 Izmir, Turkey}
\affiliation{$^4$Department of Physics, Presidency University, Kolkata, India}

\date{\today}
\begin{abstract}
We study the dynamic critical behavior of the two-dimensional random-bond three-state Potts model using large-scale Monte Carlo simulations with the Wolff and Swendsen--Wang cluster algorithms. At the disorder-dependent critical temperature, we compute integrated and exponential autocorrelation times and extract the dynamic critical exponent $z$ via finite-size scaling. Critical slowing down is significantly weakened by bond randomness, with the dynamic exponent decreasing from $z \approx 0.54$ in the pure system to $z \approx 0.24$ at strong disorder. From the finite-size scaling of the Wolff cluster size, we obtain 
$\gamma/\nu \approx 1.73$, where $\gamma$ and $\nu$ are the susceptibility 
and correlation-length exponents, respectively, consistent with the 
universality class of the two-dimensional three-state Potts model. These results indicate that bond randomness strongly affects the critical 
dynamics but leaves the underlying static universality class unchanged. Furthermore, both static and dynamic observables, including the specific heat and the autocorrelation times, show a lack of self-averaging.
\end{abstract}
\maketitle

\section{Introduction}
\label{sec:introduction}
Universality refers to the remarkable observation that physically distinct systems can exhibit identical critical behavior in the vicinity of a second-order phase transition, irrespective of their microscopic details~\cite{Stanley, Yeomans}. Systems belonging to the same universality class share identical sets of critical exponents, which describe the scaling behavior of physical observables near criticality. In this regard, spin models constitute a powerful and versatile theoretical framework for identifying and classifying universality classes~\cite{Binder, Newman, Binder2}. Specifically, the Ising model~\cite{Ising} and $q$-state Potts models~\cite{Potts} are widely employed in the study of critical phenomena in both pure and disordered spin systems. By allowing $q$ discrete spin states, the $q$-state Potts model displays a broad spectrum of critical behaviors that are not present in the Ising case~\cite{{Wu1982}}.

Understanding phase transitions and critical phenomena in regimes that are difficult or inaccessible through analytical approaches is largely enabled by Monte Carlo (MC) simulations~\cite{Binder, Newman, Binder2}. For MC simulations to operate efficiently, it is essential to understand their dynamical properties, which determine the computational performance of the algorithm near criticality. Near a second-order phase transition, critical slowing down emerges as correlations become long-ranged, which strongly affects the efficiency of numerical simulations. For systems undergoing second-order phase transitions, critical slowing down is characterized by the dynamic critical exponent $z$, which governs the scaling of the relaxation time $\tau$ with system size~\cite{Newman},
\begin{equation}
    \tau \sim L^z,
\end{equation}
where $L$ denotes the linear dimension of the system. In MC simulations of lattice models, large system sizes are required to reduce finite-size effects and statistical uncertainties. However, critical slowing down severely limits the accessible sizes, especially near criticality. For local single-spin-flip dynamics~\cite{Metropolis}, this effect is particularly pronounced: in the two-dimensional Ising model~\cite{Nightingale} and the Potts model~\cite{Tobochnik, Arcangelis, Tang}, the dynamic exponent typically takes values around $z \approx 2$. In contrast, cluster-update algorithms such as the Swendsen--Wang~\cite{Swendsen} and Wolff~\cite{Wolff1} methods significantly reduce critical slowing down, leading to substantially smaller dynamic exponents and offering a more efficient framework for exploring critical behavior~\cite{Newman, Wolff2}.

Disordered systems provide a central framework for understanding critical phenomena in diverse contexts, including quantum spin models~\cite{Sandvik, Sandvik2, Kanbur} and nonequilibrium phase transitions~\cite{Acharyya, Vatansever1, Vatansever2, Vasilopoulos}.
In spin models, randomness is typically introduced through site dilution~\cite{Fahnle, Kim, Ballesteros, Selke98, Martins, Hasenbusch1}, bond dilution~\cite{Andreichenko92, Hasenbusch3, Zhong, Kole}, or bond randomness~\cite{Cho, Fytas, Wang, Hasenbusch2} to mimic the imperfections that are inevitable in realistic materials~\cite{Young}.
In addition to its influence on static critical properties, such randomness can also affect the efficiency and dynamical behavior of numerical algorithms used to explore criticality. In this respect, spin systems with disorder provide an important testing ground for determining the response of algorithmic dynamics to randomness.

Several studies on spin models have shown that incorporating disorder can significantly influence the performance of Monte Carlo algorithms. In the site-diluted simple-cubic Ising model, it has been reported that both the autocorrelation time and the dynamic critical exponent of the Swendsen–Wang and Wolff algorithms decrease as the concentration of magnetic sites is reduced~\cite{Hennecke, Ivaneyko}. For the two-dimensional bond-diluted Ising model, algorithm-dependent behavior has been observed: while the Wolff algorithm exhibits significantly longer correlation times than in the pure Ising case, the Swendsen–Wang algorithm displays shorter correlation times and a reduced dynamic exponent~\cite{Kole}. More recently, a detailed study of cluster-update dynamics in the two-dimensional random-bond Ising model demonstrated that bond disorder substantially reduces the autocorrelation time and suppresses critical slowing down~\cite{Kanbur2024}.

While substantial progress has been made in the study of dynamic critical phenomena, investigations of disordered spin systems have so far focused predominantly on the Ising model. For models with multiple spin degrees of freedom, such as the three-state Potts model, the role of randomness in the dynamical performance of cluster algorithms remains largely unexplored. We note that dynamic properties of the random-bond Potts model have previously been examined for the $q=8$ case, where bond randomness was shown to convert the first-order transition of the pure system into a second-order one and to significantly modify the relaxation behavior~\cite{Chen1997}. However, a systematic investigation of the effect of bond randomness on the dynamic critical scaling of Potts models that already exhibit continuous transitions in the pure limit is still lacking. The three-state Potts model, a well-established model in the theory of critical phenomena, therefore provides a natural setting to address this aspect. In this paper, we investigate the dynamic properties of the Wolff and Swendsen–Wang cluster algorithms in the random-bond three-state Potts model. By analyzing autocorrelation times of several thermodynamic observables over a broad range of bond-randomness strengths and system sizes, we perform a detailed finite-size scaling analysis to determine the dynamic critical exponents. Our results indicate that increasing bond randomness leads to a reduction in both autocorrelation times and dynamic critical exponents and an associated improvement in algorithmic efficiency.

The structure of the paper is as follows. Section~\ref{sec:models} describes the model, the simulation details, and the procedure used to determine the dynamic 
critical exponents. The numerical results are presented and discussed in Section~\ref{sec:results}, and the conclusions are given in Section~\ref{sec:conclusions}.

\section{Model and Simulation Details}
\label{sec:models}
\subsection{Random-Bond Potts Model}
\noindent The random-bond $q$-state Potts model on a 2D square lattice is described by the Hamiltonian
\begin{equation}
    \mathcal{H} = - \sum_{\langle ij \rangle} J_{ij} \, \delta_{\sigma_i, \sigma_j},
    \label{hamiltonian}
\end{equation}
where the spin variables take the values $\sigma_i = 1, 2, \ldots, q$, 
and $\delta$ denotes the Kronecker delta function. The summation runs 
over all nearest-neighbor pairs of spins. The ferromagnetic exchange 
interactions $J_{ij} > 0$ are randomly selected from two positive 
couplings according to a bimodal probability distribution,
\begin{equation}
\mathcal{P}(J_{ij}) =
p\,\delta(J_{ij}-J_1)
+
(1-p)\,\delta(J_{ij}-J_2)
\label{distribution}
\end{equation}
where \( p \) denotes the probability of selecting the coupling \( J_1 \), while \( (1-p) \) corresponds to the probability of selecting \( J_2 \).
We set the Boltzmann constant to $k_B = 1$ and impose the constraint 
$J_1 + J_2 = 2$ with $J_1 > J_2$. The disorder strength is defined as 
$r = J_2/J_1$~\cite{Fytas,Vatansever1,Kanbur}. Throughout this work, we consider the values $r = 1.0$ (pure case), $r = 0.6/1.4$, $r = 0.5/1.5$, $r = 0.4/1.6$, and $r = 0.2/1.8$.

In the present study, we consider only the self-dual case with $p = 0.5$~\cite{Wu}, for which the exact critical temperature is given by~\cite{Kinzel}
\begin{equation}
\label{eq:potts_Tc}
\left( e^{J_1/T_c} - 1 \right)
\left( e^{J_2/T_c} - 1 \right) = q .
\end{equation}
We focus on the case $q = 3$, for which the model 
undergoes a continuous (second-order) phase transition. For the pure case $r=1$ and the disorder strengths
$r=0.6/1.4$, $0.5/1.5$, $0.4/1.6$, and $0.2/1.8$,
the corresponding critical temperatures obtained from
Eq.~\ref{eq:potts_Tc} are
$T_c=0.995$, $0.946$, $0.917$, $0.877$, and $0.754$,
respectively.

\subsection{Methods}
To investigate the critical dynamics of the random-bond Potts model, we employ both the Wolff and Swendsen--Wang (SW) cluster algorithms.  In the Wolff algorithm, a single cluster is constructed and flipped at each Monte Carlo step according to the following procedure:
\begin{enumerate}
    \item A lattice site is randomly selected as a seed.
    \item Nearest neighbors of the seed that are in the same Potts state 
    are added to the cluster with probability 
    $P_{\mathrm{add}} = 1 - e^{-J_{ij}/T}$.
    \item The procedure is iteratively applied to newly added spins 
    until no further spins can be incorporated into the cluster.
    \item All spins in the completed cluster are reassigned to a new 
    Potts state, chosen uniformly from the remaining $q-1$ states.
\end{enumerate}

In the SW algorithm, clusters are constructed using the same bond-activation probability $P_{\mathrm{add}}$. After all clusters are identified, each cluster is updated independently: it changes its spin state with probability $(q-1)/q$, in which case a new state is selected uniformly from the remaining $q-1$ Potts states, while it remains unchanged with probability 
$1/q$~\cite{Binder,Newman, Baillie}.

For each value of the disorder parameter $r$, we generated $10\,000$ independent bond realizations on square lattices with linear sizes $L = 16,\,24,\,32,\,48,\,64,\,96,\,128,\,192,\,256,\,384,$ employing periodic boundary conditions in all directions. 
For every realization, the system was first equilibrated over $10\,000$ Monte Carlo steps (MCS), followed by an additional $50\,000$ MCS during which thermal averages were accumulated. Simulations were initiated from a fully ordered configuration (with a specified value of spin variables in all lattice sites) to accelerate convergence toward equilibrium.

In order to analyze the dynamical properties of the system, we compute the normalized time autocorrelation function for a generic observable $A$, which in the present work corresponds to the energy $E$ of the system, and is defined as
\begin{equation}
     C_A(t)= 
     \frac{\left< A(0) A(t)\right>- \left< A\right>^2}
     {\left< A^2\right>-\left< A\right>^2}.
\end{equation}
Here, $\left<\cdots\right>$ denotes averaging over Monte Carlo time, 
and $t$ is measured in Monte Carlo steps corresponding to one full 
lattice sweep in the cluster-update scheme.

From $C_A(t)$, the integrated autocorrelation time is calculated as
\begin{equation}
    \label{eq:int}
    \tau_{A,\mathrm{int}} = \frac{1}{2} + \sum_{t=1}^{\infty} C_A(t).
\end{equation}

At sufficiently large times, the autocorrelation function follows an 
exponential decay form~\cite{jankeMonteCarloSimulations2012},
\begin{equation}
    \label{eq:exp}
     C_A(t) \sim e^{-t/\tau_{A,\mathrm{exp}}},
\end{equation}
which defines the exponential autocorrelation time 
$\tau_{A,\mathrm{exp}}$ associated with the observable $A$.

In practical simulations, the direct summation in 
Eq.~(\ref{eq:int}) becomes increasingly noisy at long times. 
To control systematic deviations, the summation is truncated at 
$t_{\mathrm{max}}$ once the self-consistency criterion 
$t_{\mathrm{max}} \ge 6\,\tau_{A,\mathrm{int}}$ is fulfilled~\cite{jankeMonteCarloSimulations2012}. 
The remaining tail contribution is then estimated using the exponential 
decay form~\cite{jankeApplicationMulticanonicalMultigrid1995},
\begin{equation}
   \tau_{A,\mathrm{int}}(t_{\mathrm{max}}) =
   \tau_{A,\mathrm{int}} -
   a\,\frac{e^{-t_{\mathrm{max}}/\tau_{A,\mathrm{exp}}}}
   {e^{1/\tau_{A,\mathrm{exp}}}-1},
\end{equation}
where the constant $a$ is determined from the exponential behavior in 
Eq.~(\ref{eq:exp}).

The observable examined in this work is the internal energy per site, 
$E=\left<\mathcal{H}\right>/N$ with $N$ 
being the total number of spins. The dynamic critical exponents 
$z_{\mathrm{exp}}$ and $z_{\mathrm{int}}$ quantify the critical 
slowing-down near $T_c$. These exponents are extracted for each disorder 
strength and for both cluster algorithms through the finite-size scaling 
of the corresponding autocorrelation times at criticality.

To ensure a proper comparison between the two cluster algorithms, 
Monte Carlo time in the Wolff algorithm is rescaled by the factor 
$f=\langle \Gamma \rangle/N$, where $\langle \Gamma \rangle$ denotes the average 
cluster size. For the finite-size scaling analysis, the quenched disorder average of 
the autocorrelation times is computed. Using the notation
$\tau_{A,\mathrm{T}}$ ($\mathrm{T}=\mathrm{int,exp}$), the rescaled
autocorrelation times are defined as
$\tau_{A,\mathrm{T}}=[f][\tau_{A,\mathrm{T}}]$,
where square brackets denote disorder averages,
in accordance with the procedure described in
Ref.~\cite{jankeNonselfaveragingAutocorrelationsPotts2000a}. 
The corresponding dynamical critical exponents $z_{A,\mathrm{T}}$ are 
then extracted from the scaling relation 
$\tau_{A,\mathrm{T}} \sim L^{z_{A,\mathrm{T}}}$ at criticality.

The statistical uncertainties of both the exponential and integrated 
autocorrelation times are evaluated using the jackknife method, 
after first grouping the data into 100 bins through a rebinning procedure. 
The dynamic critical exponents are then extracted from a weighted 
least-squares analysis performed over the data corresponding to 
different system sizes.

\section{Results and discussions}
\label{sec:results}
\begin{figure}[h!]
\includegraphics[scale=1.2]{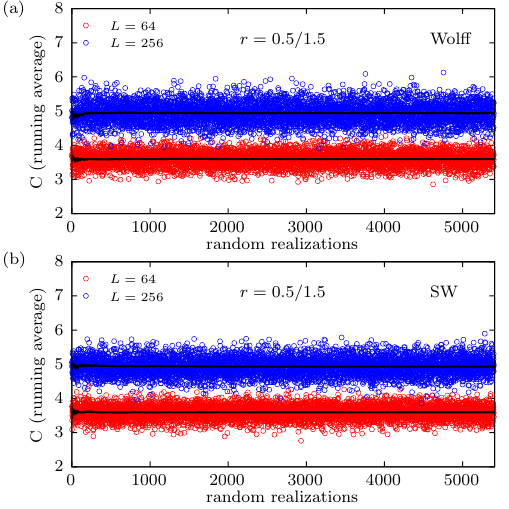}
\caption{ Running averages of the specific heat $C$ as a function of the number of random-bond realizations at bond randomness $r=0.5/1.5$. Results are shown for system sizes $L=64$ and $L=256$ using (a) the Wolff algorithm and (b) the SW algorithm. }\label{RunningAverage}
\end{figure}

Before discussing the physical results, it is important to verify that the disorder realizations used in our simulations are statistically adequate. Fig.~\ref{RunningAverage} shows the running averages of the specific heat as a function of the number of disorder realizations for $r=0.5/1.5$, and system sizes $L=64$ and $L=256$, obtained using the Wolff and SW algorithms. For both system sizes and algorithms, the running averages converge to stable plateau values after a sufficient number of samples. This confirms that the disorder averaging employed in this work is statistically well controlled and ensures reliable estimates.

We next investigate the dynamical behavior of the SW and Wolff algorithms at the disorder-dependent critical temperature obtained from Eq.~\ref{eq:potts_Tc}. The integrated and exponential autocorrelation times are summarized in Table~\ref{tbl:corr_time}. For the pure limit ($r=1$), the measured autocorrelation times are consistent with the results reported in Ref.~\cite{Baillie} for the two-dimensional $q=3$ Potts model. As expected, $\tau_{E,\exp} \geq \tau_{E,\mathrm{int}}$, in accordance with the general relation between the exponential and integrated autocorrelation times~\cite{Sokal}. When bond randomness is introduced, both $\tau_{E,\mathrm{int}}$ and $\tau_{E,\mathrm{exp}}$ decrease as the disorder parameter $r$ is reduced. For strong disorder ($r=0.2/1.8$), the autocorrelation times are significantly shorter than in the pure case, which indicates faster decorrelation of configurations.

\begin{figure}[h!]
\includegraphics[scale=1.0]{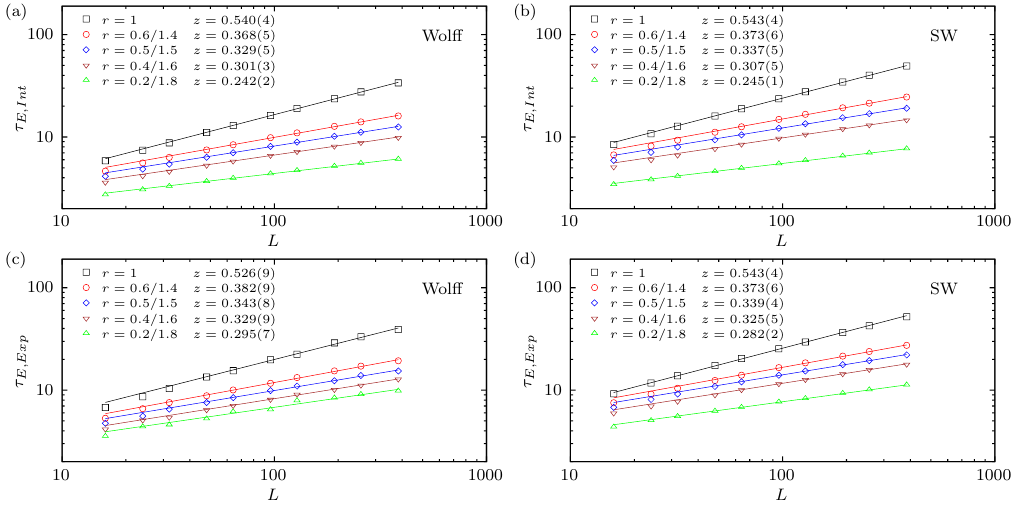}
\caption{Log-log plots of the integrated and exponential autocorrelation times of the energy, $\tau_{E,\mathrm{int}}$ and $\tau_{E,\mathrm{exp}}$, as functions of the system size $L$ for different bond-randomness strengths $r$. Panels (a) and (c) show results obtained using the Wolff algorithm, while panels (b) and (d) correspond to the SW algorithm. The straight lines represent least-squares fits to the scaling form
$\tau \sim L^{z}$, from which the dynamic critical exponent $z$ is extracted.}\label{tau_vs_L}
\end{figure}

\begin{figure}[h!]
\includegraphics[scale=1.0]{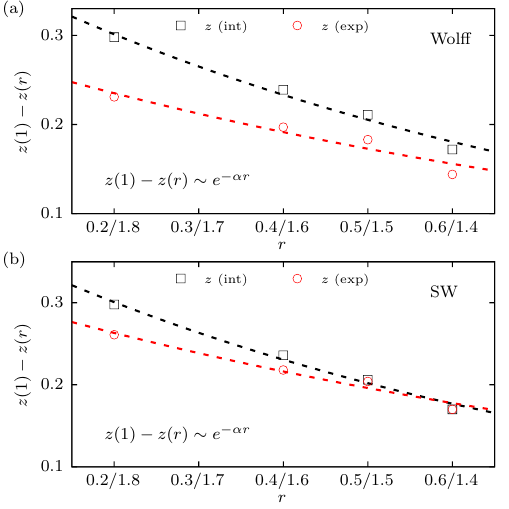}
\caption{The difference \( z(1)-z(r) \) as a function of the bond-randomness strength \( r \) for the (a) Wolff and (b) SW cluster algorithms. Here, \( z(r) \) denotes the dynamic critical exponent corresponding to the disorder strength \( r \). The dashed lines represent least-squares fits to the exponential form \( z(1)-z(r)\sim e^{-\alpha r} \).}\label{exp}
\end{figure}

The dynamic critical exponent $z$ is obtained from the finite-size scaling relation $\tau \sim L^{z}$, as illustrated in Fig.~\ref{tau_vs_L}. For the case $r=1$, the integrated autocorrelation exponents are
$z_{\mathrm{int}}=0.540(4)$ and $0.543(4)$ for the Wolff and SW algorithms,
respectively. The corresponding exponents obtained from the exponential
autocorrelation times are $z_{\mathrm{exp}}=0.526(9)$ for the Wolff algorithm
and $0.543(4)$ for the SW algorithm. These values are in good agreement with
those reported in Ref.~\cite{Baillie}. For $r=0.2/1.8$, the exponent decreases to values close to $z \approx 0.24$, with intermediate disorder strengths showing a smooth crossover between the pure and strongly disordered cases. This monotonic reduction of $z$ indicates that bond randomness suppresses the growth of relaxation times with system size, thereby weakening the critical slowing down. Consequently, both the autocorrelation times and their scaling with system size decrease as the disorder strength increases. Also, the dynamic exponent of the cluster algorithms exhibits a dependence on the disorder strength, such that the deviation of $z(r)$ from its pure value $z(1)$ follows an approximate exponential form, $z(1)-z(r) \sim e^{-\alpha r}$, with $\alpha>0$, as shown in Fig.~\ref{exp}. A similar trend was reported for the random-bond Ising model~\cite{Kanbur2024}.

We also observe that the dynamic exponents of the SW and Wolff algorithms remain very similar for all values of $r$, whereas their autocorrelation times differ noticeably.  The Wolff algorithm consistently yields smaller $\tau$ values than the SW algorithm. This difference originates from the update mechanism: in the SW scheme, multiple clusters are generated and partially flipped within a sweep, whereas the Wolff algorithm flips a single cluster at each step, leading to more efficient decorrelation. In addition, because only a single cluster is constructed during each update, the seed spin is statistically more likely to belong to a larger cluster on average, which can further enhance the efficiency of the Wolff algorithm. It is worth noting that the comparatively larger dynamic exponent observed in the pure two-dimensional three-state Potts model, relative to the Ising case (for which $z \approx 0.25$ in cluster algorithms), may arise from differences in the underlying critical fluctuations of the two models. Such differences can affect cluster formation and decorrelation dynamics, resulting in distinct values of the dynamic exponent.

For comparison, we also consider our previous study of the random-bond two-dimensional Ising model~\cite{Kanbur2024}. In the Ising model, increasing disorder strength systematically reduces both the autocorrelation times and the dynamic exponent, showing an effective softening of critical slowing down. In particular, at disorder ratio $r=0.5/1.5$, $z \approx 0.20$ was obtained for both the SW and Wolff algorithms~\cite{Kanbur2024}. Even at comparable disorder strengths, the dynamic exponents obtained for the Potts model remain substantially larger than those of the Ising model. This difference may be associated with the stronger critical fluctuations and cluster correlations of the Potts model, which can influence the dynamic scaling behavior even in the presence of quenched disorder.

\begin{figure}[h!]
\includegraphics[scale=1.2]{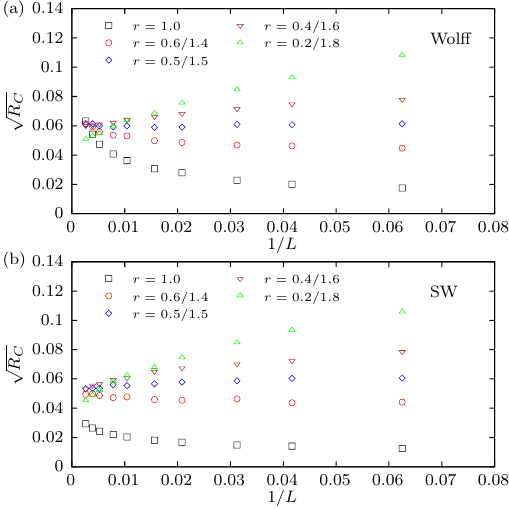}
\caption{Square root of the relative variance of the specific heat, shown as $\sqrt{R_C}$, as a function of $1/L$
for different bond-randomness strengths $r$.
Panels (a) and (b) correspond to results obtained using
the Wolff and SW algorithms, respectively.}\label{NormalizedVariance}
\end{figure}

\begin{figure}[h!]
\includegraphics[scale=1.2]{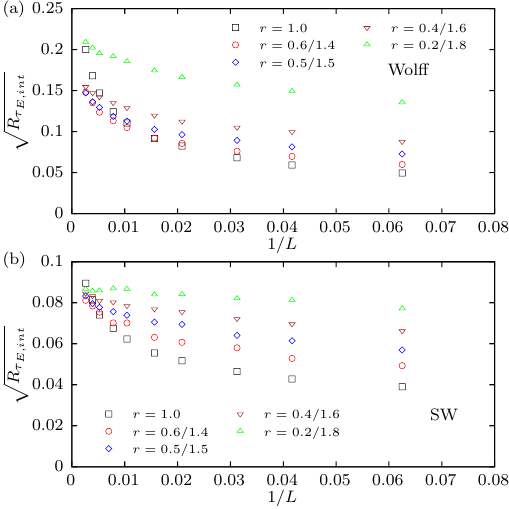}
\caption{Square root of the relative variance of the integrated energy autocorrelation time,
$\sqrt{R_{\tau_{E,\mathrm{int}}}}$, plotted as a function of $1/L$
at the disorder-dependent critical temperature for several disorder strengths $r$ for
(a) Wolff cluster algorithm (b) SW cluster algorithm.}\label{NormalizedVariance_tau}
\end{figure}

In disordered systems, an extensive thermodynamic observable depends on the particular realization of quenched disorder. As a result, measurements fluctuate from sample to sample, a phenomenon known as non-self-averaging. Self-averaging of a thermodynamic observable $A$ can be quantified through its normalized variance 
$R_A = \left([A^2] - [A]^2\right)/[A]^2$, 
where $[\cdots]$ denotes averaging over disorder realizations~\cite{wisemanSelfAveragingDistributionPseudoCritical1998a}.
If $R_A \to 0$ as $N \to \infty$, the observable is self-averaging, 
whereas a finite limiting value indicates non-self-averaging. In the latter case, extensive disorder averaging is required to obtain reliable estimates. Self-averaging in disordered spin systems has mainly been examined through static observables. Weak or absent self-averaging has been reported in several disordered spin systems, including the bond-disordered Ashkin-Teller and four-state Potts models~\cite{Wiseman2}, where magnetization and susceptibility were found to be non-self-averaging, whereas energy-related observables exhibited weak self-averaging, as well as the two-dimensional site-diluted Ising model~\cite{Tomita} and the random-bond triangular Ising model~\cite{Theodorakis12}.

Fig.~\ref{NormalizedVariance} displays the square root of the normalized variance of the specific heat as a function of the inverse system size $1/L$ for different disorder strengths, obtained using both cluster algorithms. It is worth noting that the results for the pure case ($r=1$) are primarily associated with normalized run-to-run fluctuations and finite-size effects at criticality rather than quenched-disorder-induced sample-to-sample fluctuations.
For all values of $r$, $\sqrt{R_C}$ does not vanish in the thermodynamic limit but instead approaches a finite value, demonstrating the absence of self-averaging in the specific heat. Dynamic quantities provide a complementary perspective. 
In particular, the normalized variance of the integrated energy autocorrelation time characterizes disorder-induced sample-to-sample fluctuations~\cite{jankeNonselfaveragingAutocorrelationsPotts2000a}. 
Fig.~\ref{NormalizedVariance_tau} shows $\sqrt{R_{\tau_{E_{\mathrm{int}}}}}$ as a function of the inverse system size $1/L$ for several disorder strengths, again for both cluster-update schemes. For all cases considered, $\sqrt{R_{\tau_{E_{\mathrm{int}}}}}$ exhibits a tendency toward saturation at large system sizes, which implies that relative fluctuations remain finite even in large systems. The combined static and dynamic evidence consistently demonstrates non-self-averaging in the random-bond three-state Potts model. 
A similar lack of self-averaging was reported for the random-bond Ising model~\cite{Kanbur2024}. The agreement between the two cluster algorithms further suggests that this behavior is not an artifact of the update dynamics.

In addition to the dynamical analysis, the Wolff cluster algorithm employed in our simulations provides information on static critical behavior through its cluster representation. Within the Fortuin-Kasteleyn representation, the mean size of the clusters flipped during the Wolff update is directly proportional to the magnetic susceptibility. In particular, for temperatures $T \ge T_c$, the susceptibility can be expressed as
\begin{equation}
\chi = \beta \langle \Gamma \rangle,
\end{equation}
where $\beta = 1/(k_B T)$. 

\begin{figure}[h!]
\includegraphics[scale=1.2]{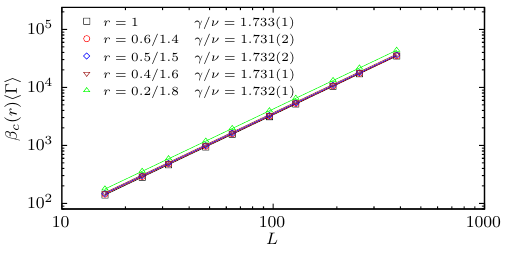}
\caption{Finite-size scaling behavior of $\beta_c(r)\langle \Gamma \rangle$ 
for different bond-randomness strengths $r$, where $\langle \Gamma \rangle$
denotes the average cluster size measured at the critical temperature.
The straight lines represent power-law fits to the scaling form
$\beta_c(r)\langle \Gamma \rangle \sim L^{\gamma/\nu}$.
}\label{Gamma_nu}
\end{figure}

According to finite-size scaling theory, the magnetic susceptibility at criticality follows the power-law behavior
\begin{equation}
\chi(L) \sim L^{\gamma/\nu},
\end{equation}
which allows the exponent ratio $\gamma/\nu$ to be extracted from the system-size dependence of $\beta_c(r)\langle \Gamma \rangle$. As shown in Fig.~\ref{Gamma_nu}, we determine $\gamma/\nu$ for all investigated bond-randomness strengths by performing least-squares fits to the corresponding power-law form at the disorder-dependent critical temperature $T_c(r)$. 
Our analysis yields $\gamma/\nu \approx 1.73$ within numerical uncertainties for both the pure system and the random-bond cases, in agreement with the exact value $\gamma/\nu \approx 1.733$ of the two-dimensional three-state Potts universality class~\cite{Kim1996, Picco1996}. Within error bars, the exponent ratio remains essentially unchanged as the strength of bond randomness increases. This behavior is consistent with the scenario of weak universality reported for the random-bond three-state Potts model, where individual critical exponents may vary with disorder strength, while exponent ratios such as $\gamma/\nu$ and $\beta/\nu$ remain invariant~\cite{Kim1996, Picco1996}. Our findings, therefore, indicate that, although disorder significantly influences the dynamical properties of the cluster algorithms, the static scaling relations given in the exponent ratios are preserved in the parameter regime considered.

\begingroup
\squeezetable
\begin{table*}[t]
\vspace{-2.5cm}
\caption{Energy autocorrelation times 
($\tau_{E,\mathrm{exp}}$ and $\tau_{E,\mathrm{int}}$)
at $T_c$ for different system sizes $L$ and disorder strengths $r$,
obtained using the SW and Wolff algorithms.}
\label{tbl:corr_time}
\begin{ruledtabular}
\begin{tabular}{cc cc cc}
\multicolumn{2}{c}{} & \multicolumn{2}{c}{SW} & \multicolumn{2}{c}{Wolff} \\

$L$ & $r$
& $\tau_{E,exp}$ & $\tau_{E,int}$
& \multicolumn{1}{c}{$\tau_{E,exp}$}
& \multicolumn{1}{c}{$\tau_{E,int}$} \\
\addlinespace[3pt]
\hline
\addlinespace[3pt]

16  & 0.2/1.8 & 4.35(1) & 3.43(1) & 3.53(2) & 2.72(1) \\
    & 0.4/1.6 & 5.98(1) & 5.11(1) & 4.18(1) & 3.62(1) \\
    & 0.5/1.5 & 6.79(1) & 5.92(1) & 4.74(1) & 4.13(1) \\
    & 0.6/1.4 & 7.53(1) & 6.69(1) & 5.31(1) & 4.64(1) \\
    & 1.0 & 9.23(1) & 8.44(1) & 6.77(1) & 5.86(1) \\
\addlinespace[4pt]

24  & 0.2/1.8 & 5.04(1) & 3.82(1) & 4.38(4) & 3.06(1) \\
    & 0.4/1.6 & 7.01(1) & 5.99(1) & 5.11(1) & 4.19(1) \\
    & 0.5/1.5 & 8.12(1) & 7.08(1) & 5.59(1) & 4.87(1) \\
    & 0.6/1.4 & 9.14(1) & 8.15(1) & 6.59(1) & 5.58(1) \\
    & 1.0 & 11.72(1) & 10.75(1) & 8.63(1) & 7.43(1) \\
\addlinespace[4pt]

32  & 0.2/1.8 & 5.52(1) & 4.13(1) & 4.57(4) & 3.30(1) \\
    & 0.4/1.6 & 7.80(1) & 6.66(1) & 5.45(1) & 4.63(1) \\
    & 0.5/1.5 & 9.16(1) & 8.00(1) & 6.51(1) & 5.44(1) \\
    & 0.6/1.4 & 10.43(1) & 9.31(1) & 7.56(1) & 6.33(1) \\
    & 1.0 & 13.83(1) & 12.72(1) & 10.36(1) & 8.77(1) \\
\addlinespace[4pt]

48  & 0.2/1.8 & 6.23(1) & 4.59(1) & 5.23(6) & 3.68(1) \\
    & 0.4/1.6 & 9.00(1) & 7.70(1) & 6.44(1) & 5.30(1) \\
    & 0.5/1.5 & 10.86(1) & 9.40(1) & 7.58(1) & 6.34(1) \\
    & 0.6/1.4 & 12.43(1) & 11.14(1) & 8.82(1) & 7.48(1) \\
    & 1.0 & 17.36(1) & 16.07(1) & 13.44(1) & 11.04(1) \\
\addlinespace[4pt]

64  & 0.2/1.8 & 6.78(1) & 4.94(1) & 6.12(16) & 3.95(1) \\
    & 0.4/1.6 & 10.11(1) & 8.52(1) & 7.10(2) & 5.82(1) \\
    & 0.5/1.5 & 12.02(1) & 10.49(1) & 8.45(1) & 7.03(1) \\
    & 0.6/1.4 & 14.00(1) & 12.60(1) & 10.04(2) & 8.41(1) \\
    & 1.0 & 20.33(1) & 18.91(1) & 15.51(2) & 12.98(1) \\
\addlinespace[4pt]

96  & 0.2/1.8 & 7.64(1) & 5.47(1) & 6.43(12) & 4.37(1) \\
    & 0.4/1.6 & 11.56(1) & 9.73(1) & 8.14(3) & 6.62(1) \\
    & 0.5/1.5 & 13.98(1) & 12.17(1) & 9.89(3) & 8.12(1) \\
    & 0.6/1.4 & 16.59(1) & 14.86(2) & 11.66(1) & 9.86(1) \\
    & 1.0 & 25.34(1) & 23.68(2) & 19.84(3) & 16.22(1) \\
\addlinespace[4pt]

128 & 0.2/1.8 & 8.27(1) & 5.88(1) & 7.87(32) & 4.70(1) \\
    & 0.4/1.6 & 12.72(1) & 10.66(1) & 9.12(3) & 7.22(1) \\
    & 0.5/1.5 & 15.40(1) & 13.46(1) & 10.97(3) & 8.92(1) \\
    & 0.6/1.4 & 18.47(1) & 16.65(2) & 13.27(7) & 10.98(1) \\
    & 1.0 & 29.53(1) & 27.68(3) & 22.45(4) & 18.96(1) \\
\addlinespace[4pt]

192 & 0.2/1.8 & 9.29(1) & 6.50(1) & 8.37(32) & 5.17(2) \\
    & 0.4/1.6 & 14.45(1) & 12.07(1) & 10.16(3) & 8.14(1) \\
    & 0.5/1.5 & 17.76(1) & 15.43(2) & 12.34(3) & 10.18(1) \\
    & 0.6/1.4 & 21.50(1) & 19.29(2) & 15.41(3) & 12.71(1) \\
    & 1.0 & 36.62(1) & 34.42(4) & 28.94(9) & 23.70(2) \\
\addlinespace[4pt]

256 & 0.2/1.8 & 10.05(1) & 6.97(1) & 9.07(33) & 5.54(2) \\
    & 0.4/1.6 & 15.95(1) & 13.16(1) & 11.19(7) & 8.85(1) \\
    & 0.5/1.5 & 19.46(1) & 16.88(2) & 13.92(6) & 11.14(1) \\
    & 0.6/1.4 & 23.84(1) & 21.40(2) & 17.16(5) & 14.06(1) \\
    & 1.0 & 42.46(1) & 40.06(5) & 33.30(13) & 27.60(1) \\
\addlinespace[4pt]

384 & 0.2/1.8 & 11.22(1) & 7.66(1) & 9.80(33) & 6.07(2) \\
    & 0.4/1.6 & 17.93(1) & 14.72(2) & 12.84(14) & 9.92(1) \\
    & 0.5/1.5 & 22.10(1) & 19.14(2) & 15.38(8) & 12.53(1) \\
    & 0.6/1.4 & 27.38(1) & 24.58(3) & 19.35(5) & 16.06(1) \\
    & 1.0 & 52.20(1) & 49.47(6) & 39.05(29) & 33.80(1) \\

\end{tabular}
\end{ruledtabular} 
\end{table*}
\endgroup

\section{Conclusions}
\label{sec:conclusions}

We have investigated the dynamic critical behavior of the two-dimensional random-bond three-state Potts model using large-scale Monte Carlo simulations with the Wolff and Swendsen--Wang cluster algorithms. 
By analyzing both integrated and exponential autocorrelation times at the disorder-dependent critical temperature, we extracted the dynamic critical exponent over a broad range of bond-randomness strengths.

Our results show that increasing bond randomness systematically reduces the autocorrelation times and the dynamic critical exponent, leading to a pronounced weakening of critical slowing down. 
The dynamic exponent decreases smoothly from its pure value to significantly smaller values at strong disorder, and its deviation from the disorder-free case follows an approximate exponential dependence on the disorder parameter.

Despite the substantial modification of the dynamical properties, the static exponent ratio $\gamma/\nu$, obtained from the finite-size scaling of the Wolff cluster size, remains consistent with the universality class of the pure three-state Potts model within numerical accuracy. 
Furthermore, both static and dynamic observables exhibit non-self-averaging behavior, indicating persistent sample-to-sample fluctuations in the presence of quenched bond randomness.

These findings indicate that quenched bond randomness significantly alters dynamic scaling while preserving static universality in the parameter regime considered. 
Future studies may examine whether similar behavior persists for other values of $q$ or for different types of disorder distributions.

\begin{acknowledgments}
The numerical calculations reported in this paper were performed
at T\"{U}B\.{I}TAK ULAKBIM (Turkish agency), High Performance and
Grid Computing Center (TRUBA Resources).
\end{acknowledgments}

\end{document}